# New Closed Virial Equation of State for Hard-Sphere Fluids

Jianxiang Tian [1, 2, 4], Yuanxing Gui [2], Angel Mulero [3]

[1]Shandong Provincial Key Laboratory of Laser Polarization and Information Technology

Department of Physics, Qufu Normal University, Qufu, 273165, P. R. China

[2]Department of Physics, Dalian University of Technology, Dalian, 116024, P. R. China

[3]Department of Applied Physics, University of Extremadura, Badajoz 06072, Spain

[4]Corresponding author, E-mail address: jxtian@dlut.edu.cn

**Abstract**

A new closed virial equation of state of hard-sphere fluids is proposed which reproduces the calculated or estimated values of the first sixteen virial coefficients at the same time as giving very good accuracy when compared with computer simulation data for the compressibility factor over the entire fluid range, and having a pole at the correct closest packing density.

**Introduction**

The study of the hard-sphere (HS) fluid has long been a subject of great interest. The state-of-the-art of the topic up to the year 2008 was reviewed in a recent book[1], and three new studies[2-4] have been published very recently proposing new analytical expressions for the HS equation of state, all based on closed virial forms. They were checked for their reproduction of both computer simulation data for the compressibility factor and the estimates of the eleventh to sixteenth virial coefficients given by Clisby and McCoy[5].

These three proposals have as yet not been compared with each other, and there are now also new and very accurate computer simulation data available reported by Bannerman et al.[4] The following is a brief overview of the current situation:

(i) From the results given in both our previous[2] (see Table 4 therein) and the Bannerman et al. [4] (Fig. 3 therein) studies, it is clear that the most accurate currently available HS EOS at reduced densities below 1 is that denominated KLM1 in Ref. 2, proposed by Kolafa et al.[6]. This is based on the use of the values of the first six virial coefficients and then on the calculation of five adjustable coefficients to fit the authors' own computer simulation data. This success nevertheless does not carry over to metastable densities for which the KLM1 EOS deviates even more than the well-known and simple Carnahan-Starling EOS[7]. Indeed, we have shown[2] that the Padé approximant [4/5], CM45, proposed by Clisby and McCoy[5] is the most accurate EOS for densities from 0.95 to 1.09 (computer simulation data from Kolafa and coworkers[6, 8] were used as reference). This high density range corresponds to metastable states, since the onset of freezing is at a density of 0.9435. As a conclusion, the CM45 EOS can be considered as the most accurate over the whole density range.

(ii) With respect to the reproduction of the estimates of the eleventh to sixteenth virial coefficients as given by Clisby and McCoy[5], we have shown[2] that the EOS denominated AEM(-5,2), based on the use of the asymptotic expansion method (AEM) [9] and which takes the calculated first ten virial coefficients as reference, is the only one of those that we analyzed that gives these coefficients within their estimation errors.

(iii) Of the EOSs we have considered above, only the recent proposals of Hu and Yu (HY)[3] and Bannerman *et al.* (BLW)[4] have a pole at the correct closest packing density, $\sqrt{2}$, which is commonly considered as a limiting reference point with physical meaning when describing the behaviour of hard spheres through equations of state. Nevertheless, this requirement is not always achieved. Thus, for instance, the recent KLM1 and CM45 EOSs have a pole at $6/\pi \approx 1.91$, and AEM(-5,2) at around 1.77, which are excessively high values. Some other expressions consider also a random close packing density, at which the fluid branch of the phase diagram must have another pole.[1]

(iv) Neither the two proposed HY EOSs[3] nor the two proposed BLW EOSs[4] were included in our previous study[2]. Furthermore, neither the AEM nor the HY EOSs have been compared with the new computer simulation data obtained very recently by Bannerman *et al.*[4]. In any case, the results presented in the aforementioned three references seem to indicate that there is currently no EOS that can reproduce very accurately the compressibility factor data at metastable densities at the same time as having a pole at the proper closest packing density, and giving values for the highest virial coefficients in good agreement with those estimated by Clisby and McCoy[5] inside their calculated deviations.

In light of the above results, it was clearly desirable to appropriately compare the behaviour of the HY, BLW, and AEM EOSs with each other and with the newly available computer simulation data, and also to try to find a new closed-form virial EOS that can improve the predictions of the compressibility factor data at metastable densities at the same time as maintaining some of the good predictions of previous EOSs and containing the estimated virial coefficients from the eleventh to the sixteenth one.

This paper is organized as follows. In Sec. 2 we briefly provide some details about previous proposals. Section 3 describes our new proposal, presents the new results, and compares them with those of previous proposals. And finally, Sec. 4 gives the conclusions.

## 2. Recent closed-form virial EOSs

As is well-known, the virial series expansion of the compressibility factor, $Z$, is:

$$Z = \frac{P}{\rho k_B T} = 1 + \sum_{n=2}^{\infty} B_n y^{n-1} \qquad (1)$$

where $P$ is the pressure, $\rho$ the reduced density, $T$ the temperature, $k_B$ Boltzmann's constant, and the packing fraction, $y$, is defined in HS fluids as $y = \pi\rho/6$, the closest packing value being $y_c \approx 0.7405$. For the HS case, only the first four $B_n$ values can be calculated analytically, whereas numerical values for $B_5$ to $B_{10}$ and estimated values for $B_{11}$ to $B_{16}$ have been obtained by Clisby and McCoy[5]. These estimated values, together with their percentage deviations, are given in the first column of Table 1. Despite some of them are given with deviations around 3-4%, these values can be considered as a valuable reference in order to build new equations of state or to check if some previous equations lead to the same or similar values.

Different kinds of approximants have been used in order to obtain a closed-form expression from Eq. (1) (see a summary in Ref. 1). As noted above, the most accurate EOSs currently available are those of Kolafa et al.[6] and Clisby-McCoy[5], even though they are not so accurate at the highest metastable densities. For the highest virial coefficients, they give values which do not agree[2] with the values given by Clisby and McCoy (listed in Table 1). Although, of course, these are not serious inconvenient, it is still possible to propose new expressions, which give better accuracy at metastable densities as the same time that giving the first sixteen virial coefficients in good agreement with the currently available estimated values.

Those previous EOSs were constructed using only the first few virial coefficients, but in the last two years five new EOSs have been proposed considering closed analytical virial expressions including the first ten or even higher virial coefficients as inputs [2-4]. We shall briefly describe these new EOSs.

Hu and Yu (HY) [3] have proposed different Padé-type approximants in order to reproduce the values of the virial coefficients higher than the second. In particular, they showed that it is enough to use a [3/3] Padé approximant, which requires the calculated values for the first nine virial coefficients. Since they proposed new values for $B_7$ and $B_{10}$, the $B$ values obtained do not agree with those given by Clisby and McCoy [5], as can be seen in Table 1. In particular, there are deviations greater than 6% for $B_{14}$ to $B_{16}$.

At the same time, they proposed two similar EOSs. The first, which we shall call HY1, is Eq. (8) in Ref. 3. It is based on a sum of the infinite sequence from order higher than a given $m$ value. The second, which we shall call HY2, is Eq. (11) in Ref. 3. It has a separate term using the second virial coefficient. Both of them allow the closest packing fraction to be explicitly included. They also both require the values of the first eighteen virial coefficients (see Table 6 in Ref. 3) to be considered (*i.e.*, $m = 18$) in order to obtain low average absolute deviations (AAD) with respect to computer simulation data for $Z$. As can be seen in Table 2, the HY2 EOS gives lower AAD values than does HY1 over the whole density range. One also observes in Table 2 that this EOS gives slightly lower AAD values than the CM54 Padé approximant [5] in both the stable and the metastable density ranges. As shown in Fig. 1, this improvement of the HY2 EOS is due to its excellent behaviour compared with most other EOSs at densities from 0.85 to 0.94, but it gives high deviations at lower densities.

Almost simultaneously with the HY paper, we proposed a new AEM EOS, Eq. (10) in Ref. 2, denominated AEM(-5,2), which was constructed using the calculated values for the first ten virial coefficients and that can also reproduce the estimated values of $B_{11}$ to $B_{16}$, as one observes in Table 1. Unfortunately, this EOS has a pole at the too high value of $y = 0.9262$. As can be seen in Table 2, it is less accurate than other EOSs at the stable densities, and of similar accuracy to most of the others in the metastable range. Hence its main advantage is that, by using only ten virial coefficients, it is capable of reproducing all the sixteen virial coefficients locating at the error region.

Even more recently, Bannerman et al.[4] have considered two different approaches to reproducing the behaviour of the highest known virial coefficients. The first takes $(B_n y_c^{n-1} - B_{n-1} y_c^n)$ to be approximately the constant value 0.68219 for $n=10$ to $n=12$. Then a new EOS was proposed which we shall call WC1, and is Eq. (6) in Ref. 4. This equation was firstly proposed by Prof. Woodcock himself.[10] It includes the calculated values for the first ten virial coefficients and the aforementioned constant value. As one sees in Table 1, we found that the $B_{11}$ to $B_{13}$ values are slightly overestimated when compared with those of Clisby and McCoy, and that the higher virial coefficients are accurately reproduced within the estimated errors.

The second proposal of Bannerman et al.[4] is based on the fact that the differences $(B_n y_c^{n-1} - B_{n-1} y_c^n)$ for $n = 8$ to 10 can be approximated by an exponential function which does not include any adjustable parameter. So the virial sum is extended to the first nine virial coefficients and then the other terms are represented by the exponential function. The new EOS, which we shall call WC2, is Eq. (7) in Ref. 4. As can be seen in Table 1, the values of $B_{11}$, $B_{13}$, and $B_{14}$ agree with the estimates of Clisby and McCoy[5], whereas the $B_{15}$ and $B_{16}$ values are clearly lower than the estimated ones. This disagreement should be checked again when new calculated virial coefficients are known. As one observes in Table 2 and Fig. 1, the improvement of WC2 EOS over WC1 (both include the correct value of the closest packing fraction) is clear in the stable density range.

Consequently, the improvement that currently can be made would be to construct a new EOS that could reproduce simultaneously the calculated or estimated values of the first sixteen virial coefficients and of the compressibility factor, and that has the correct closest packing density limit. We shall here describe the construction of a new EOS which is based on the study of the behaviour of the known highest virial coefficients, on the use of one adjustable coefficient to reproduce the compressibility factor data, and on taking into account the limiting behaviour of both the virial coefficients and the compressibility factor at the closest packing density.

## 3. New Closed Virial Equation of State for Hard Spheres

In order to propose a new closed virial equation, we consider first the relationship between the limiting behaviour of the highest virial coefficients for hard spheres and the location of the first pole. In particular, Yelash et al.[11] have shown that:

$$\lim_{n \to \infty} (B_n / B_{n+1}) = y_{pole} \tag{2a}$$

In order to build a new EOS with a pole at the closest packing fraction, we assume that $y_{pole} = y_c \approx 0.7405$. Thus, for a high $n$ value, we are going to use the simple relationship:

$$(B_n / B_{n+1}) = y_c. \tag{2b}$$

Using equation (2b), our new closed virial EOS is based on the separation of the virial expansion, Eq. (1), into three parts. We consider separately the first $n-1$ terms of the virial expansion, then the terms from $n$ to $m$ which can be taken as linear with respect to their order, and then the remaining terms which are taken to be appropriate to give the correct pole in the EOS. In particular, the HS EOS is written as:

$$Z = Z_T + Z_L + Z_I \tag{3}$$

where

$$Z_T = 1 + B_2 y + B_3 y^2 + \cdots + B_n y^{n-1} \tag{4}$$

$$Z_L = \sum_{i=n}^{m}(c_1+c_2 i)y^i \tag{5}$$

$$Z_I = \sum_{j=m+1}^{\infty} c_0 y^j / y_c^j \tag{6}$$

$Z_T$ is the virial series truncated to *(n-1)*th order. $Z_L$ is used to account for the virial terms for which the coefficients $B_{n+1}$ to $B_{m+1}$ are linear with their order ever observed by Woodcock [10] for $B_9$ to $B_{12}$, with $c_1$ and $c_2$ being adjustable coefficients giving that linear relationship. $Z_I$ is designed to satisfy the limiting behaviour of higher virial coefficients proofed by Yelash et al.,[11] with $c_0 = B_{m+2} y_c^{m+1}$ being regarded as an adjustable coefficient because of the lack of the knowledge of very high virial coefficients.

The closed form of Eq. (3) can be written as follows:

$$\begin{aligned}Z = &1+B_2 y + B_3 y^2 + \cdots + B_n y^{n-1} + \\ &\frac{(c_1+c_2 n)y^n + (c_2-c_1-c_2 n)y^{n+1} + (-c_2-c_1-c_2 m)y^{m+1} + (c_1+c_2 m)y^{m+2}}{(1-y)^2} \\ &+ \frac{c_0 y^{m+1}}{y_c^{m+1}(1-y/y_c)}\end{aligned} \tag{7}$$

where appropriate values of *m, $c_1$, $c_2$,* and $c_0$ must be chosen in order to give adequate values for the higher virial coefficients and also for the Z values over the entire density range. Equation (7) has been successfully used for description of properties of hard disk fluids. [12]

As noted above, only the first ten $B_i$ values have been calculated. Hence, in order to obtain the proper expression for $Z_L$, we studied the behaviour of the $B_{11}$ to $B_{16}$ estimates of Clisby and McCoy [5] versus their order. We found that only the values from $B_{13}$ to $B_{16}$ can be represented by a linear relationship, with $c_1$=-209.7845 and $c_2$=32.615 being the corresponding parameters. Hence, one has to take *n* = 12. The virial coefficients obtained are given in Table 1, and obviously are very close to those of Ref. 5.

To find the appropriate value of *m*, we determined the AADs, as defined in Table 2, for values of *m* greater than 12. Values greater than 30 were found not to influence the results, and the lowest AADs were obtained for $m=30$, so that our choice was $m=30$. As can be seen in Table 2 and Fig. 1, $Z_T+Z_L$ ($n=12$, $m=30$) behaves similarly to WC1 in the stable density range, similarly to HY2 in the metastable range, and similarly to AEM(-5,2) in the two ranges. The AAD for the entire density range is 0.29%, and the pole is located at $y = 1$, so that this new EOS does not present any clear advantage over the others discussed above. The third term, $Z_I$, is therefore needed in order to achieve improvements over the previous results.

We therefore considered the $Z_I$ term, obtaining the appropriate $c_0$ value from a fit of Eq. (7) to the computer simulation results in the metastable region. The lowest AAD2 value in the high density region was 0.12% for $c_0 = 393.3499$ and $m=30$, this being almost five times lower than the AAD2 given by the most accurate EOS in this range, CM45, as one sees from Table 2. As a result, the AAD3 for the entire density range was only 0.06%, which is around four times lower than that corresponding to the CM45 EOS. Unfortunately, this improvement in the metastable region meant a slight increase in the deviations in the density range from 0.9 to 0.94. This is seen in Fig. 1, which shows that this new EOS behaves similarly to CM45. In any case, this new EOS, Eq. (7), is the only one presently available that can reproduce the presently known virial coefficients at the same time as having the correct closest packing density limit and giving an AAD below 0.12% in the stable, metastable, and entire density regions for which computer simulation values of $Z$ are available.

## 4. Conclusions

Several recent proposals for the equation of state of hard-sphere fluids have been considered and analyzed in order to reproduce the calculated first ten virial coefficients and the estimated values published by Clisby and McCoy[5] for the 11th to the 16th, at the same time as having good accuracy in the calculation of the compressibility factor and a pole at the correct closest packing density.

All the equations of state analyzed or proposed were based on the virial expansion, and were considered together and then compared for the first time in this study. In particular, we have shown that the asymptotic expansion method is the only method of those analyzed here that can give an equation of state reproducing the estimated values of the first sixteen virial coefficients from knowledge of only the values of the first ten. All the other methods considered here to accelerate the convergence of the series, and to yield adequate values of the virial coefficients, require knowledge of at least the values of the first twelve.

A new analytical expression for the equation of state of hard-sphere fluids has been proposed. It comprises three terms: (i) a truncated virial series including the values of the first ten virial coefficients; (ii) a second term in which the virial coefficients from the 13th to the 30th are taken to be linear with their order (the two parameters needed for the linear relationship were determined using the estimated values of the 13th to the 16th virial coefficients); and (iii) an asymptotic term that takes the limiting behaviour of virial coefficients into account and thus includes a pole at the correct closest packing density. This last term requires a fit to the values of the compressibility factor given by computer simulations – in the present proposal this is done with only one adjustable parameter.

We then showed that this new equation of state is the only one that can reproduce the estimated values of the first sixteen virial coefficients at the same time as giving only small deviations for the compressibility factor values over the entire density range, and also having a pole at the correct closest packing density.

Finally we would stress that the proposals considered here are based on what are presently considered the best estimates of the 11th to the 16th virial coefficients. While these coefficients may be calculated more accurately in the near future, the resulting values could maintain their linearity with respect to their order to certain extent. As long as this is the case, the new values can be incorporated into the proposed EOS with no difficulty. As has been shown in the present study, other methods may require major modifications and/or the inclusion of further adjustable coefficients.

**Acknowledgements**

The National Natural Science Foundation of China under Grant No. 10804061, the Natural Science Foundation of Shandong Province under Grant No. Y2006A06, and the foundations of QFNU and DUT provided support for this work (J.T. and Y.G.).

**References**

[1] Mulero, A. *Theory and Simulation of Hard-Sphere Fluids and Related Systems*, Lect. Notes Phys. 753, Springer, Berlin Heidelberg, 2008. Chapter 3 (and references therein).

[2] Tian, J. X.; Gui, Y. X.; Mulero, A. *Phys. Chem. Chem. Phys.*, **2009**, *11*, 11213-11218.

[3] Hu, J. W.; Yu, Y. X. *Phys. Chem. Chem. Phys.*, **2009**, *11*, 9382-9390.

[4] Bannerman, M. N.; Lue, L.; Woodcock, L. V. *J. Chem. Phys.*, **2010**, *132*, 084507.

[5] Clisby, N.; B. M. McCoy, *J. Stat. Phys.*, **2006**, *122*, 15-57.

[6] Kolafa, J.; Labik, S.; Malijevsky, A. *Phys. Chem. Chem. Phys.*, **2004**, *6*, 2335-2340.

[7] Carnahan, N. F. ; Starling, K. E. *J. Chem. Phys.*, **1969**, *51*, 635-636.

[8] Kolafa, J. *Phys. Chem. Chem. Phys.*, **2006**, *8*, 464-468.

[9] Khanpour, M.; Parsafar, G. A. *Chem. Phys.*, **2007**, *333*, 208-213.

[10] Woodcock, L. arxiv:0801.4846v3, **2008**.

[11] Yelash, L. V. ; Kraska T.; Deiters, U. K. *J. Chem. Phys.*, **1999**, *110*, 3079-3084.

[12] Tian, J. X.; Gui Y. X.; Mulero, A. *Phys. Chem. Chem. Phys.*, **2010**, accepted, DOI:10.1039/C0CP00476F.

**Table Captions**

**Table 1** Predicted virial coefficients of HS fluids. The percentage deviation with respect to Ref. 5 is given in parentheses. Coefficients in bold italic type were used explicitly to construct the EOS.

**Table 2** AADs (%) of the results of EOSs for HS fluids over three different density ranges. AAD1 = stable range, for densities from 0.1 to 0.94 (Ref. 4). AAD2 = metastable range, for densities from 0.95 to 1.09 (Refs. 4, 6, 8). AAD3 = entire density range. The lowest values are in bold. Details of the EOSs are given in the main text. NRS = no real solution.

**Figure Captions**

**Fig. 1** Differences between Z values from computer simulation [4] and from EOSs, $\Delta Z = Z_{sim} - Z_{eos}$, versus density, for the highest density stable region.

Table 1 Predicted virial coefficients of HS fluids. The percentage deviation with respect to Ref. 5 is given in parentheses. Coefficients in bold italic type were used explicitly to construct the EOS.

| Virials | Estimated values (Ref.5) | Z(-5,2) (Ref.2) | WC1 (Ref.4) | WC2 (Ref.4) | HY (Ref.3) | Eq. (7) |
|---|---|---|---|---|---|---|
| $B_{11}$ | 127.93 (0.82%) | 127.58 (0.27%) | 129.08 (0.90%) | 128.69 (0.59%) | 126.22 (1.34%) | *__127.93__* |
| $B_{12}$ | 152.67 (0.28%) | 152.61 (0.04%) | 155.73 (2.00%) | 154.43 (1.15%) | 149.37 (2.16%) | *__152.67__* |
| $B_{13}$ | 181.19 (0.93%) | 180.82 (0.20%) | 185.21 (2.22%) | 182.28 (0.60%) | 174.49 (3.70%) | 181.60 (0.23%) |
| $B_{14}$ | 214.75 (3.1%) | 212.56 (1.02%) | 216.22 (0.68%) | 210.62 (1.92%) | 201.59 (6.13%) | 214.21 (0.25%) |
| $B_{15}$ | 246.96 (1.1%) | 248.21 (0.51%) | 246.21 (0.30%) | 236.40 (4.28%) | 230.69 (6.59%) | 246.83 (0.05%) |
| $B_{16}$ | 279.17 (3.9%) | 288.19 (3.23%) | 270.67 (3.04%) | 254.37 (8.88%) | 261.77 (6.23%) | 279.44 (0.10%) |

**Table 2** AADs (%) of the results of EOSs for HS fluids over three different density ranges. AAD1 = stable range, for densities from 0.1 to 0.94 (Ref. 4). AAD2 = metastable range, for densities from 0.95 to 1.09 (Refs. 4, 6, 8). AAD3 = entire density range. The lowest values are in bold. Details of the EOSs are given in the main text. NRS = no real solution.

| EOS | Ref. | $y_{pole}$ | AAD1 (%) | AAD2 (%) | AAD3 (%) |
|---|---|---|---|---|---|
| CS | 7 | 1 | 0.159 | 0.72 | 0.40 |
| CM45 | | 0.8580 | | | |
| | 5 | 1.0983 | 0.023 | 0.56 | 0.25 |
| | | 1.5790 | | | |
| CM54 | 5 | NRS | 0.004 | 0.76 | 0.33 |
| KLM1 | 6 | 1 | **0.001** | 0.60 | 0.26 |
| AEM(-5,2) | 2 | 0.9262 | 0.013 | 0.64 | 0.28 |
| HY1 | 3 | 0.7405 | 0.006 | 0.76 | 0.33 |
| HY2 | 3 | 0.7405 | 0.003 | 0.67 | 0.29 |
| WC1 | 4 | 0.7405 | 0.014 | 0.79 | 0.35 |
| WC2 | 4 | 0.7405 | 0.007 | 0.88 | 0.38 |
| $Z_T+Z_L$ | this work | 1 | 0.014 | 0.66 | 0.29 |
| $Z_T+Z_L+Z_I$ | this work | 1; 0.7405 | 0.018 | **0.11** | **0.06** |

**Fig. 1** Differences between Z values from computer simulation [4] and from EOSs, ΔZ =Zsim-Zeos, versus density, for the highest density stable region.

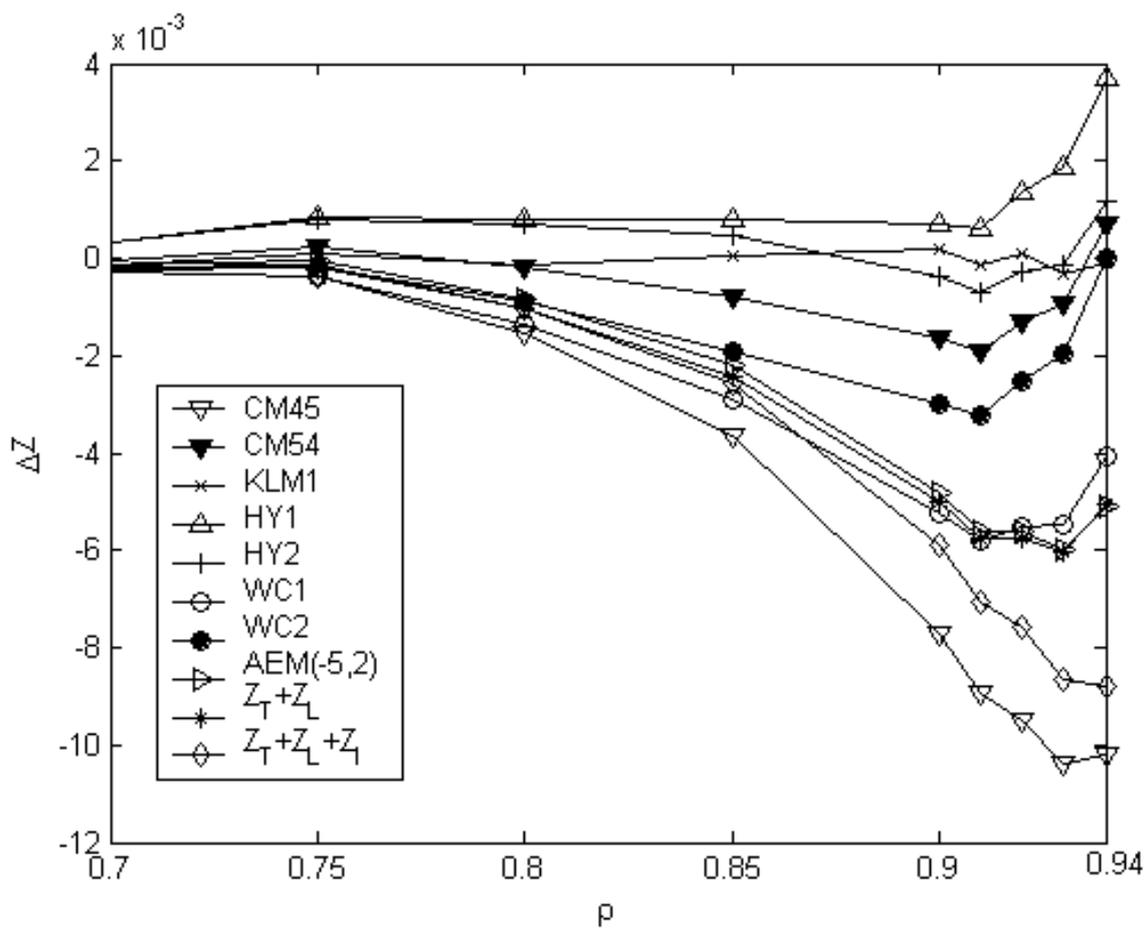

**Table of Contents Image**

$$Z = 1 + B_2 y + B_3 y^2 + \cdots + B_n y^{n-1} +$$
$$\frac{(c_1 + c_2 n) y^n + (c_2 - c_1 - c_2 n) y^{n+1} + (-c_2 - c_1 - c_2 m) y^{m+1} + (c_1 + c_2 m) y^{m+2}}{(1-y)^2}$$
$$+ \frac{c_0 y^{m+1}}{y_c^{m+1}(1 - y/y_c)}$$

New equation of state for hard-sphere fluids from virial coefficients and three adjustable parameters